\documentstyle [12pt]{article}
\makeatletter \oddsidemargin  0in \evensidemargin 0in
\textwidth16cm \RequirePackage[dvips]{graphicx} \textheight 20.5cm
\newcommand{\singlespacing}{\let\CS=\@currsize\renewcommand{\baselinestretch}{1.5}\tiny\CS}
\makeatletter \oddsidemargin  -.1in \evensidemargin -.1in
\newcommand{\doublespacing}{\let\CS=\@currsize\renewcommand{\baselinestretch}{1.35}\tiny\CS}

\def\@citex[#1]#2{\if@filesw\immediate\write\@auxout{\string\citation{#2}}\fi
  \def\@citea{}\@cite{\@for\@citeb:=#2\do
    {\@citea\def\@citea{,\linebreak[0]\hskip0pt plus .2em}%
      \@ifundefined{b@\@citeb}%
    {{\bf ?}\@warning{Citation `\@citeb' on page \thepage\space undefined}}%
      \hbox{\csname b@\@citeb\endcsname}}}{#1}}

\newtheorem{rule-def}[theorem]{Rule}
\begin{document}
\title{\bf Single qubit , two qubit gates and no
signalling principle}\author{I.Chakrabarty
$^{1,2}$\thanks{Corresponding author:
E-Mail-indranilc@indiainfo.com }\\
$^1$ Heritage Institute of Technology,Kolkata-107,West Bengal,India\\
$^2$ Bengal Engineering and Science University, Howrah, West
Bengal, India }
\date{}
\maketitle{}
\begin{abstract}
In this work we investigate that whether one can construct single
and two qubit gates for arbitrary quantum states from the
principle of no signalling. We considered the problem for Pauli
gates, Hadamard gate, C-Not gate.
\end{abstract}
\section{Introduction}
In quantum information theory, there are many information
processing protocols or operations which can not be carried out
perfectly for an unknown qubit. This may be probably due to the
linear structure or may be due to the unitary evolution in
quantum mechanics. Regardless of their origin, these impossible
operations are making quantum information processing more
restricted than it's classical counterpart. On the other hand
these restrictions on many quantum information processing tasks
are making quantum information more secure. Enlisting of these
operations started from the landmark paper of Wootters and Zurek,
where 'no-cloning' theorem has been stated [1]. This theorem tells
us that one cannot clone a single quantum. Later it was also
shown by Pati and Braunstein that we cannot delete either of the
two quantum states when we are provided with two identical
quantum states at our input port [2]. Even after the no-deletion
theorem, many other operations like 'self replication', 'partial
erasure', 'splitting' proved as impossible operations in quantum
domain [3,4,5]. These 'no-go'theorems come under the broad heading
of 'General impossible operations' [6]. Researches are carried
out to see how these no go theorems in quantum information theory
are consistent with various principles of quantum mechanics. One
of such principle is the principle of no signalling. It tells us
that if two distant parties Alice and Bob, share an entangled
state, neither Alice nor Bob cannot send signal faster than the
speed of light to the other party by doing some local operation
on their own subsystems. It had been already seen that if one
assumes these impossible operations to be valid physical
processes, one can have a super luminal communication between two
distant parties sharing an entangled state
[7,8,9,10,12,13,14,15,16,17,18]. These results also guarantees
impossibility of such operations from the no signalling principle.
Here in this work we will address the question that whether one
can construct the single qubit and two qubit gates for
nonorthogonal states, and we find that it is impossible to do so,
as this will violate the principle of no signalling. The entire
organization of the work is as follows: In the second section we
will discuss the existing proofs of impossibility of various
operations from the no signalling principle. In the third section
we will consider one qubit gates like Pauli gates, Hadamard gate
and will show their impossibility from the principle of no
signalling. In the fourth section we will show the same for two
qubit gates.
\section{Revisiting impossible operations and no signalling
principle:} \textbf{Cloning and no signalling:} It is a well
known fact that there exists no physical process by which one can
achieve the transformation $|\psi_i\rangle \longrightarrow
|\psi_i\rangle|\psi_i\rangle$ for a set of non orthogonal states
$\{|\psi_i\rangle\}$ [1,11]. One can easily prove that if we
assume cloning of an unknown quantum to be a feasible operation,
then one can send signals faster than the speed of light [7]. Let
two distant parties Alice and Bob share a singlet state,
\begin{eqnarray}
|X\rangle=\frac{1}{\sqrt{2}}[|\psi\rangle|\bar{\psi}\rangle-|\bar{\psi}\rangle|\psi\rangle]
\end{eqnarray}
Since the singlet state remains invariant in any arbitrary qubit
basis, then after Alice carries out a measurement on her
subsystem in any two basis, the resultant reduced density matrix
on Bob's side is $\frac{I}{2}$ . This clearly indicates that
initially under normal scenario, Bob cannot distinguish the
statistical mixtures representing his subsystem obtained as a
result of measurement carried out by Alice in two different basis.
Henceforth it is not possible for Bob to obtain information
regarding the basis on which Alice has performed her measurement.
However if Bob attaches ancilla to his qubit and perfectly clone
his qubit then the entangled state takes the form
\begin{eqnarray}
|X\rangle^C=\frac{1}{\sqrt{2}}[|\psi\rangle|\bar{\psi}\rangle|\bar{\psi}\rangle-|\bar{\psi}\rangle|\psi\rangle|\psi\rangle]
\end{eqnarray}
Now if Alice performs measurement on her qubit, on two different
basis $\{|\psi_1\rangle,|\bar{\psi_1}\rangle\}$ and
$\{|\psi_2\rangle,|\bar{\psi_2}\rangle\}$, then the reduced
density matrices describing Bob's subsystem are given by
\begin{eqnarray}
\rho_1^C=\frac{1}{2}[|\psi_1\psi_1\rangle\langle
\psi_1\psi_1|+|\bar{\psi_1}\bar{\psi_1}\rangle\langle
\bar{\psi_1}\bar{\psi_1}|]
\end{eqnarray}
\begin{eqnarray}
\rho_2^C=\frac{1}{2}[|\psi_2\psi_2\rangle\langle
\psi_2\psi_2|+|\bar{\psi_2}\bar{\psi_2}\rangle\langle
\bar{\psi_2}\bar{\psi_2}|]
\end{eqnarray}
Now Bob can easily distinguish the statistical mixture obtained
as a result of Alice's measurement and subsequently can infer on
which basis Alice has performed measurement. This clearly
indicates that super luminal signalling has taken place. Hence
forth we can conclude that perfect deterministic cloning of an
unknown quantum state is not a feasible operation.\\
\textbf{General Impossible operations and no signalling:} In this
subsection we will see that 'General Impossible Operations'[6]
which will act on the tensor product of an unknown quantum state
and blank state at the input port to produce the original state
along with a function of the original state at the output port is
not feasible in the quantum world from the no signalling
principle [9]. Suppose there is a singlet state consisting of two
particles shared by two distant parties Alice and Bob. The state
is given by
\begin{eqnarray}
|\chi\rangle_{12}=\frac{1}{\sqrt{2}}(|0\rangle|1\rangle-|1\rangle|0\rangle)\nonumber\\
=\frac{1}{\sqrt{2}}(|\psi\rangle|\bar{\psi}\rangle-|\psi\rangle|\bar{\psi}\rangle)
\end{eqnarray}
where $\{|\psi\rangle,|\bar{\psi}\rangle\}$ are mutually
orthogonal spin states or in other words they are mutually
orthogonal polarizations in case of photon particles. Alice is in
possession of the first particle
and Bob is in possession of the second particle.\\
No-signalling principle states that if one distant partner (say,
Alice) measures her particle in any one of the two basis namely
$\{|0\rangle,|1\rangle\}$ and
$\{|\psi\rangle,|\bar{\psi}\rangle\}$, then measurement outcome
of the other party (say, Bob) will remain invariant. At this
point one might ask an interesting question: Is there any
possibility for Bob to know the basis in which Alice measured her
qubit, if he applies the operations defined as 'General
Impossible operations'[6] on his
qubit.\\
 Let us consider a situation where Bob is in possession of
a hypothetical machine whose action in two different basis
$\{|0\rangle,|1\rangle\}$ and
$\{|\psi\rangle,|\bar{\psi}\rangle\}$ is defined by the
transformation,
\begin{eqnarray}
|i\rangle|\Sigma\rangle\longrightarrow|i\rangle|F(i)\rangle (i=0,1)\\
|j\rangle|\Sigma\rangle\longrightarrow|j\rangle|F(j)\rangle
(j=\psi,\bar{\psi})
\end{eqnarray}
where $|\Sigma\rangle$ is the ancilla state attached by Bob .
These set of
transformations was first introduced by Pati in [6].\\
After the application of the transformation defined in (6-7) by
Bob on his particle, the singlet state takes the form
\begin{eqnarray}
|\chi\rangle|\Sigma\rangle=\frac{1}{\sqrt{2}}(|0\rangle|1\rangle|F(1)\rangle-|1\rangle|0\rangle|F(0)\rangle)
\nonumber\\=\frac{1}{\sqrt{2}}(|\psi\rangle|\bar{\psi}\rangle|F(\bar{\psi})\rangle-|\bar{\psi}\rangle|\psi\rangle|F(\psi)\rangle)
\end{eqnarray}
Now Alice can measure her particle in two different basis
$\{|0\rangle,|1\rangle\}$ and
$\{|\psi\rangle,|\bar{\psi}\rangle\}$ , then the reduced density
matrices describing Bob's subsystem are given by,
\begin{eqnarray}
\rho_1=\frac{1}{2}[|1\rangle\langle 1 |\otimes|F(1)\rangle\langle
F(1) |+|0\rangle\langle 0|\otimes |F(0)\rangle\langle F(0) |]
\end{eqnarray}
\begin{eqnarray}
\rho_2=\frac{1}{2}[|\bar{\psi}\rangle\langle \bar{\psi}
|\otimes|F(\bar{\psi})\rangle\langle F(\bar{\psi})
|+|\psi\rangle\langle \psi|\otimes |F(\psi)\rangle\langle F(\psi)
|]
\end{eqnarray}
Since the statistical mixture in (9) and (10) are different, so
this would have allow Bob to distinguish the basis in which Alice
has performed the measurement and this lead to super luminal
signalling. But this is not possible from the principle of
'no-signalling', so we arrive at a contradiction. Hence, we
conclude from the principle of no-signalling that the
transformation defined in (6-7) is not possible in the quantum
world. Here one can easily see that cloning of a quantum state is
a special case of general impossible operations and impossibility
of these operations defined in (6-7)from no signalling principle
once again proves the impossibility of cloning.\\
\textbf{Deletion and no signalling:} In reference [8], Pati and
Braunstein showed that the deletion of a arbitrary quantum state
implies signalling. They have considered a situation where two
distant parties Alice and Bob shared two singlet states. The
combined state of the system in arbitrary qubit basis
$\{|\psi\rangle,|\bar{\psi}\rangle\}$is given by,
\begin{eqnarray}
|\chi\rangle_{12}|\chi\rangle_{34}=\frac{1}{2}[|\psi\rangle_1|\psi\rangle_2|\bar{\psi}\rangle_3|\bar{\psi}\rangle_4+|\bar{\psi}\rangle_1|\bar{\psi}\rangle_2|\psi\rangle_3|\psi\rangle_4
\nonumber\\-|\bar{\psi}\rangle_1|\psi\rangle_2|\psi\rangle_3|\bar{\psi}\rangle_4-|\psi\rangle_1|\bar{\psi}\rangle_2|\bar{\psi}\rangle_3|\psi\rangle_4]
\end{eqnarray}
Now if Alice measures her particles in any qubit basis, and if
she doesn't convey her measurement result to Bob, then Bob's
particles are in completely random mixture i.e
$\rho_{24}=\frac{I}{2}\otimes \frac{I}{2}$ \\
But suppose Bob has a quantum deleting machine which can delete
arbitrary quantum state. The action of the deleting machine can
be described by,
\begin{eqnarray}
|\psi\rangle|\psi\rangle|A\rangle\longrightarrow|\psi\rangle|\Sigma\rangle|A_{\psi}\rangle\nonumber\\
|\bar{\psi}\rangle|\bar{\psi}\rangle|A\rangle\longrightarrow|\bar{\psi}\rangle|\Sigma\rangle|A_{\bar{\psi}}\rangle\nonumber\\
|\psi\rangle|\bar{\psi}\rangle|A\rangle\longrightarrow|\phi'\rangle\nonumber\\
|\bar{\psi}\rangle|\psi\rangle|A\rangle\longrightarrow|\phi''\rangle
\end{eqnarray}
Now if Bob applies the above described deleting machine on his
particles, the combined system (11) no longer remains in the
previous form. Now if Alice performs measurement on either of two
choices of basis states $\{|0\rangle,|1\rangle\}$ and
$\{|\psi\rangle,|\bar{\psi}\rangle\}$, then the resultant reduced
density matrices describing Bob's subsystem for two different
measurement will be different.
\begin{eqnarray}
\rho(0)=\frac{1}{4}[I_2\otimes|\Sigma\rangle\langle
\Sigma|+\rho'(0)+\rho''(0)]\nonumber\\
\rho(\theta)=\frac{1}{4}[I_2\otimes|\Sigma\rangle\langle
\Sigma|+\rho'(\theta)+\rho''(\theta)]
\end{eqnarray}
If Alice measures her particle in $\{|0\rangle,|1\rangle\}$ basis
, then Bob's particle will be in $\rho(0)$, however if Alice
measures her particle in  $\{|\psi\rangle,|\bar{\psi}\rangle\}$,
then Bob's particle will be represented by  $\rho(\theta)$. Thus
it is clear that the reduced density matrix describing Bob's
subsystem are no longer completely random, but depend upon the
choice of basis. Since it is not random Bob can easily
distinguish these two density matrices and can infer about the
basis on on which Alice has performed the measurement. This leads
to super luminal signalling. This leads us to contradiction to
the initial assumption that perfect deletion is possible.
\section{\textsc{Single Qubit , Two Qubit Gates And No Signalling }}
Pauli Gates and No signalling: In this subsection we will
investigate the question whether one can construct the Pauli
gates : X, Y, Z gates , for unknown qubit from the principle of
no signalling. X gate: The importance of this gate is immense in
quantum information theory. It is also known as a spin flip
operator, as it flips a known quantum state into its orthogonal
state. However one cannot construct a universal NOT (X-gate)for
arbitrary quantum state. However the largest set of states that
can be flipped by using single NOT gate is the set lying on a
great circle of the Bloch-sphere. In reference [12], authors
established this impossibility of construction of universal not
gate from the principle of no-signalling. The protocol involved
two distant parties sharing an entangled state of the form
\begin{eqnarray}
|\Psi\rangle_{AB}=\frac{1}{\sqrt{3}}[|0\rangle_A|0\rangle_B+|1\rangle_A|\psi\rangle_B+|2\rangle_A|\phi\rangle_B]
\end{eqnarray}
where Alice's system is a three dimensional Hilbert space having
$\{|0\rangle,|1\rangle,|2\rangle\}$ as basis. Bob's system
consists of three states $\{|0\rangle, |\psi\rangle,
|\phi\rangle\}$, where
\begin{eqnarray}
|\psi\rangle=a|0\rangle+b|1\rangle\nonumber\\
|\phi\rangle=c|0\rangle+d\exp(i\theta)|1\rangle
\end{eqnarray}
( where $a^2 + b^2 = c^2 + d^2 = 1; 0 < \theta, \pi; a > 0, c >
0$). Not only that Bob is in possession of hypothetical flipping
machine, whose action is defined by
\begin{eqnarray}
|0\rangle|M\rangle\longrightarrow|1\rangle|M_0\rangle\nonumber\\
|\psi\rangle|M\rangle\longrightarrow\exp(i\mu)|\bar{\psi}\rangle|M_{\psi}\rangle\nonumber\\
|\phi\rangle|M\rangle\longrightarrow\exp(i\nu)|\bar{\phi}\rangle|M_{\phi}\rangle
\end{eqnarray}
(where $\mu$ and $\nu$ are some arbitrary phases and $|M\rangle$
is the initial machine state. Initially if we trace out Bob's
qubit the reduced density matrix describing Alice's subsystem is
given by,
\begin{eqnarray}
\rho_A^I=\frac{1}{3}[I+a(|0\rangle\langle 1 |+|1\rangle\langle 0
|)+c(|0\rangle\langle 2 |+|2\rangle\langle 0 |)+\langle \psi
|\phi\rangle|1\rangle\langle 2 |+\langle \phi
|\psi\rangle|2\rangle\langle 1|]
\end{eqnarray}
Now if Bob applies the hypothetical flipping machine (16) on his
qubit the entangled state (14) will take a new form and
correspondingly the density matrix representation of Alice's
subsystem will be of the form
\begin{eqnarray}
\rho_A^F=\frac{1}{3}[I-a(\exp(-i\mu)\langle M_{\psi}
|M_{0}\rangle|0\rangle\langle 1 |+\exp(i\mu)\langle M_{0}
|M_{\psi}\rangle|1\rangle\langle 0
|)\nonumber\\-c(\exp(-i\nu)\langle M_{\phi}
|M_{0}\rangle|0\rangle\langle 2 |+\exp(i\nu)\langle M_{0}
|M_{\phi}\rangle|2\rangle\langle 0 |)+\nonumber\\\langle \psi
|\phi\rangle \exp(i[\mu-\nu])\langle M_{\phi}
|M_{\psi}\rangle|1\rangle\langle 2 |+\langle \phi
|\psi\rangle\exp(i[\nu-\mu])\langle M_{\psi}
|M_{\phi}\rangle|2\rangle\langle 1|]
\end{eqnarray}
Since the flipping operation defined in (16) is a trace
preserving quantum operation and there is no classical
communication between two distant parties, so from the principle
of no signalling one can easily conclude that the two density
matrices  and  will be identical. However a simple calculation
reveals that the expressions (17) and(18) are not identical as
long as the states are not lying on the same great circle.
Henceforth one can conclude that it is impossible to construct a
universal NOT gate from the principle of no
signalling.\\
\textbf{Y gate:} This is another single qubit gate, which cannot
be constructed for any arbitrary qubit. Here in this subsection we
will show that if we assume the construction of this gate for
arbitrary qubit, this will lead to the violation of causality.
Let us assume that two spatially separated parties Alice and Bob
share an entangled state of the form ,
\begin{eqnarray}
|X\rangle_{AB}=\frac{1}{2}[|0\rangle_A|\psi_1\rangle_B+|1\rangle_A|\bar{\psi_1}\rangle_B+|2\rangle_A|\psi_2\rangle_B+|3\rangle_A|\bar{\psi_2}\rangle_B]
\end{eqnarray}
(where A denotes Alice's qubit, while B denotes Bob's qubit). Now
if one traces out Bob's qubit , the reduced density matrix
describing Alice's subsystem will be given by,
\begin{eqnarray}
\rho_A=\frac{1}{4}[I+|0\rangle\langle 2|\langle
\psi_2|\psi_1\rangle+|0\rangle\langle 3|\langle
\bar{\psi_2}|\psi_1\rangle+|1\rangle\langle 2|\langle
\psi_2|\bar{\psi_1}\rangle+|1\rangle\langle 3|\langle
\bar{\psi_2}|\bar{\psi_1}\rangle\nonumber\\+|2\rangle\langle
0|\langle \psi_1|\psi_2\rangle+|2\rangle\langle 1|\langle
\bar{\psi_1}|\psi_2\rangle+|3\rangle\langle 0|\langle
\psi_1|\bar{\psi_2}\rangle+|3\rangle\langle 1|\langle
\bar{\psi_1}|\bar{\psi_2}\rangle]
\end{eqnarray}
Let us assume that some how Bob has constructed a hypothetical Y
Gate for non orthogonal set of qubits. The action of such a gate
is defined by,
\begin{eqnarray}
|\psi_i\rangle\longrightarrow-i|\bar{\psi_i}\rangle\nonumber\\
|\bar{\psi_i}\rangle\longrightarrow i|\psi_i\rangle
\end{eqnarray}
(where i = 1, 2).\\
Now if Bob applies the transformation (21) on his qubit, the
entangled state reduces to the form
\begin{eqnarray}
|X\rangle_{AB}^F=\frac{1}{2}[-i|0\rangle_A|\bar{\psi_1}\rangle_B+i|1\rangle_A|\psi_1\rangle_B-i|2\rangle_A|\bar{\psi_2}\rangle_B+i|3\rangle_A|\psi_2\rangle_B]
\end{eqnarray}
As a consequence the reduced density matrix representing Alice's
subsystem will be of the form
\begin{eqnarray}
\rho_A^Y=\frac{1}{4}[I+|0\rangle\langle 2|\langle
\bar{\psi_2}|\bar{\psi_1}\rangle-|0\rangle\langle 3|\langle
\psi_2|\bar{\psi_1}\rangle-|1\rangle\langle 2|\langle
\bar{\psi_2}|\psi_1\rangle+|1\rangle\langle 3|\langle
\psi_2|\psi_1\rangle\nonumber\\+|2\rangle\langle 0|\langle
\bar{\psi_1}|\psi_2\rangle-|2\rangle\langle 1|\langle
\psi_1|\bar{\psi_2}\rangle-|3\rangle\langle 0|\langle
\bar{\psi_1}|\psi_2\rangle+|3\rangle\langle 1|\langle
\psi_1|\psi_2\rangle]
\end{eqnarray}
It is clearly evident that the expressions (20) and (23) are not
identical for all sets of qubits on the Bloch sphere. However
causality demands these expressions to be equal. This is a
violation causality. So one can say that it is impossible to
construct a universal Y gate.\\
\textbf{Z gate:} In this subsection we show that it is not
possible to construct an universal Z gate by making the
construction of such a gate consistent with the principle of no
signalling. In other words if we start with a set consisting of
non orthogonal quantum states
$\{|\psi_i\rangle,|\bar{\psi_i}\rangle\}$ where (i = 1, 2), then
from the principle of no signalling one cannot achieve the
transformation
\begin{eqnarray}
|\psi_i\rangle\longrightarrow|\psi_i\rangle\nonumber\\
|\bar{\psi_i}\rangle\longrightarrow -|\bar{\psi_i}\rangle
\end{eqnarray}
In order to have a proof of the above statement, quite likely to
other proofs we consider a situation where two distant partners
are sharing an entangled state of the form
\begin{eqnarray}
|X\rangle_{AB}=\frac{1}{2}[|0\rangle_A|\psi_1\rangle_B+|1\rangle_A|\bar{\psi_1}\rangle_B+|2\rangle_A|\psi_2\rangle_B+|3\rangle_A|\bar{\psi_2}\rangle_B]
\end{eqnarray}
One can easily obtain the reduced density matrix of Alice's
system in order to have an idea of her subsystem. The reduced
density matrix describing Alice subsystem is given by
\begin{eqnarray}
\rho_A=\frac{1}{4}[I+|0\rangle\langle 2|\langle
\psi_2|\psi_1\rangle+|0\rangle\langle 3|\langle
\bar{\psi_2}|\psi_1\rangle+|1\rangle\langle 2|\langle
\psi_2|\bar{\psi_1}\rangle+|1\rangle\langle 3|\langle
\bar{\psi_2}|\bar{\psi_1}\rangle\nonumber\\+|2\rangle\langle
0|\langle \psi_1|\psi_2\rangle+|2\rangle\langle 1|\langle
\bar{\psi_1}|\psi_2\rangle+|3\rangle\langle 0|\langle
\psi_1|\bar{\psi_2}\rangle+|3\rangle\langle 1|\langle
\bar{\psi_1}|\bar{\psi_2}\rangle]
\end{eqnarray}
The no signalling principle demands that one cannot send
information with a speed faster than the speed of light. In other
words if one of the two distant partners carries out local on his
qubit, it will not change the reduced density matrix of other
party instantaneously. However we find here that if Bob applies
this gate defined by (24) on his qubit, the reduced density
matrix describing Alice's subsystem will be different from what
it was initially. The reduced density matrix describing Alice's
subsystem after Bob's application of hypothetical Z gate on his
qubit, will be of the form
\begin{eqnarray}
\rho_A^F=\frac{1}{4}[I+|0\rangle\langle 2|\langle
\psi_2|\psi_1\rangle-|0\rangle\langle 3|\langle
\bar{\psi_2}|\psi_1\rangle-|1\rangle\langle 2|\langle
\psi_2|\bar{\psi_1}\rangle+|1\rangle\langle 3|\langle
\bar{\psi_2}|\bar{\psi_1}\rangle\nonumber\\+|2\rangle\langle
0|\langle \psi_1|\psi_2\rangle-|2\rangle\langle 1|\langle
\bar{\psi_1}|\psi_2\rangle-|3\rangle\langle 0|\langle
\psi_1|\bar{\psi_2}\rangle+|3\rangle\langle 1|\langle
\bar{\psi_1}|\bar{\psi_2}\rangle]
\end{eqnarray}
It is clearly evident that equations (26) and (27) are not
identical. This clearly indicates that super luminal signalling
has taken place, which is an impossible phenomenon in principle.
So we arrive at a contradiction and conclude that, one cannot
design universal Z gate as it will violate the principle of no
signalling.\\
\textbf{Hadamard gate and No signalling principle:} This is yet
another gate which has got immense application in quantum
information theory. The interesting question is that can we
design a universal Hadamard gate. What does no signalling
principle tells us? The answer to this question is no. In
references [9,19] authors showed that construction of universal
Hadamard gate will violate no signalling principle. In this
section we put forward a proof used in those references.\\
Now, we define the Hadamard transformation for arbitrary qubit in
the following way:
\begin{eqnarray}
|\psi_i\rangle|M\rangle\longrightarrow\frac{1}{\sqrt{2}}(|\psi_i\rangle+e^{i\phi_i}|\bar{\psi_i}\rangle)|H_{\psi_i}\rangle\nonumber\\
|\bar{\psi_i}\rangle|M\rangle\longrightarrow\frac{1}{\bar{2}}(|\psi_i\rangle-e^{i\phi_i}|\bar{\psi_i}\rangle)|H_{\bar{\psi_i}}\rangle
\end{eqnarray}
(i=1,2),where $\langle \psi_i|\bar{\psi_i}\rangle = 0$.\\
The entangled state shared between two distant partners is given
by
\begin{eqnarray}
|\Psi\rangle_{AB}=\frac{1}{\sqrt{2}}[|0\rangle|\psi_1\rangle+|1\rangle|\psi_2\rangle]|M\rangle
\end{eqnarray}
Before and after the application of Hadamard transformation on
Bob's qubit, the reduced density matrices describing the Alice's
subsystem are given by,
\begin{eqnarray}
\rho_A=\frac{1}{2}[|0\rangle\langle 0|+|1\rangle\langle
1|+|0\rangle\langle 1|(\langle
\psi_2|\psi_1\rangle)+|1\rangle\langle 0|(\langle
\psi_1|\psi_2\rangle)]
\end{eqnarray}
and
\begin{eqnarray}
&&\rho_A^H=\frac{1}{4}[|0\rangle\langle 0|(2)+|1\rangle\langle
1|(2){}\nonumber\\&&+|0\rangle\langle 1|(\langle
\psi_2|\psi_1\rangle+\langle \bar{\psi_2}|\psi_1\rangle+\langle
\psi_2|\bar{\psi_1}\rangle+\langle
\bar{\psi_2}|\bar{\psi_1}\rangle)(\langle H_{\psi_2}
|H_{\psi_1}\rangle){}\nonumber\\&&+|1\rangle\langle 0|(\langle
\psi_1|\psi_2\rangle+\langle \bar{\psi_1}|\psi_2\rangle+\langle
\psi_1|\bar{\psi_2}\rangle+\langle
\bar{\psi_1}|\bar{\psi_2}\rangle)(\langle H_{\psi_1}
|H_{\psi_2}\rangle)]
\end{eqnarray}
It is clear from equations (31) and (32) that the reduced density
matrices  and are different. This implies that by designing the
perfect Hadamard gate, one can send information faster than
light, which is impossible. Hence perfect construction of
universal Hadamard gate is not possible.
\section{Two Qubit Gates and No signalling principle:}
\textbf{C-Not Gate and No signalling: }In this section we will
show that it is impossible to construct C-Not gates for a set of
non orthogonal qubits from the no signalling principle. Controlled
-Not gate is a two qubit gate which acts as a Not gate to the
second qubit (target qubit), when the first qubit (control qubit)
is set to lie in the computational basis $\{|0\rangle,|1\rangle$.
The action of this two qubit gate in the computational basis
$\{|0\rangle,|1\rangle$, is given by,
\begin{eqnarray}
|0\rangle|0\rangle\longrightarrow|0\rangle|0\rangle\nonumber\\
|0\rangle|1\rangle\longrightarrow|0\rangle|1\rangle\nonumber\\
|1\rangle|0\rangle\longrightarrow|1\rangle|1\rangle\nonumber\\
|1\rangle|1\rangle\longrightarrow|1\rangle|0\rangle
\end{eqnarray}
At this point one may ask an interesting question that if we are
provided with a set consisting of non orthogonal quantum states
$|\psi_i\rangle$ is it possible for us to construct such a gate.
Let us assume that construction of such a gates for non
orthogonal states is possible. The action of such a gate is
described by,
\begin{eqnarray}
|\psi_i\rangle|\psi_i\rangle\longrightarrow|\psi_i\rangle|\psi_i\rangle\nonumber\\
|\psi_i\rangle|\bar{\psi_i}\rangle\longrightarrow|\psi_i\rangle|\bar{\psi_i}\rangle\nonumber\\
|\bar{\psi_i}\rangle|\psi_i\rangle\longrightarrow|\bar{\psi_i}\rangle|\bar{\psi_i}\rangle\nonumber\\
|\bar{\psi_i}\rangle|\bar{\psi_i}\rangle\longrightarrow|\bar{\psi_i}\rangle|\psi_i\rangle
\end{eqnarray}
Let us consider the situation where two distant parties Alice and
Bob share an entangled state of the form,
\begin{eqnarray}
|X\rangle=\frac{1}{2}[|0\rangle_A(|\bar{\psi_1}\rangle|\psi_1\rangle)_B+|1\rangle_A(|\bar{\psi_1}\rangle|\bar{\psi_1}\rangle)_B
+|2\rangle_A(|\bar{\psi_2}\rangle|\psi_2\rangle)_B+|3\rangle_A(|\bar{\psi_2}\rangle|\bar{\psi_2}\rangle)_B]
\end{eqnarray}
where $\{|0\rangle,|1\rangle,|2\rangle,|3\rangle\} $ are the
basis vectors of the Hilbert space describing Alice's subsystem.
Now after tracing out Bob's qubit the reduced density matrix
describing Alice's subsystem is given by
\begin{eqnarray}
\rho_A=\frac{1}{4}[I+|2\rangle\langle 0 |\{\langle
\bar{\psi_1}|\bar{\psi_2}\rangle\langle
\psi_1|\psi_2\rangle\}+|3\rangle\langle 0 |\{\langle
\bar{\psi_1}|\bar{\psi_2}\rangle\langle
\psi_1|\bar{\psi_2}\rangle\}\nonumber\\+|2\rangle\langle
1|\{\langle \bar{\psi_1}|\bar{\psi_2}\rangle\langle
\bar{\psi_1}|\psi_2\rangle\} +|3\rangle\langle 1 |\{\langle
\bar{\psi_1}|\bar{\psi_2}\rangle\langle
\bar{\psi_1}|\bar{\psi_2}\rangle\}\nonumber\\+|0\rangle\langle 2
|\{\langle \bar{\psi_2}|\bar{\psi_1}\rangle\langle
\psi_2|\psi_1\rangle\}+|1\rangle\langle 2 |\{\langle
\bar{\psi_2}|\bar{\psi_1}\rangle\langle
\psi_2|\bar{\psi_1}\rangle\}\nonumber\\+|0\rangle\langle
3|\{\langle \bar{\psi_2}|\bar{\psi_1}\rangle\langle
\bar{\psi_2}|\psi_1\rangle\} +|1\rangle\langle 3|\{\langle
\bar{\psi_2}|\bar{\psi_1}\rangle\langle
\bar{\psi_2}|\bar{\psi_1}\rangle\}]
\end{eqnarray}
Now if Bob applies the C-Not gate ,defined by equation (34), on
his qubit, then the initially shared entangled state takes the
form
\begin{eqnarray}
|X\rangle^{C-Not}=\frac{1}{2}[|0\rangle_A(|\bar{\psi_1}\rangle|\bar{\psi_1}\rangle)_B+|1\rangle_A(|\bar{\psi_1}\rangle|\psi_1\rangle)_B
+|2\rangle_A(|\bar{\psi_2}\rangle|\bar{\psi_2}\rangle)_B+|3\rangle_A(|\bar{\psi_2}\rangle|\psi_2\rangle)_B]
\end{eqnarray}
Henceforth the reduced density matrix describing Bob's subsystem
is given by,
\begin{eqnarray}
\rho_A^{C-Not}=\frac{1}{4}[I+|2\rangle\langle 0 |\{\langle
\bar{\psi_1}|\bar{\psi_2}\rangle\langle
\bar{\psi_1}|\bar{\psi_2}\rangle\}+|3\rangle\langle 0 |\{\langle
\bar{\psi_1}|\bar{\psi_2}\rangle\langle
\bar{\psi_1}|\psi_2\rangle\}\nonumber\\+|2\rangle\langle
1|\{\langle \bar{\psi_1}|\bar{\psi_2}\rangle\langle
\psi_1|\bar{\psi_2}\rangle\} +|3\rangle\langle 1 |\{\langle
\bar{\psi_1}|\bar{\psi_2}\rangle\langle
\psi_1|\psi_2\rangle\}\nonumber\\+|0\rangle\langle 2 |\{\langle
\bar{\psi_2}|\bar{\psi_1}\rangle\langle
\bar{\psi_2}|\bar{\psi_1}\rangle\}+|1\rangle\langle 2 |\{\langle
\bar{\psi_2}|\bar{\psi_1}\rangle\langle
\bar{\psi_2}|\psi_1\rangle\}\nonumber\\+|0\rangle\langle
3|\{\langle \bar{\psi_2}|\bar{\psi_1}\rangle\langle
\psi_2|\bar{\psi_1}\rangle\} +|1\rangle\langle 3|\{\langle
\bar{\psi_2}|\bar{\psi_1}\rangle\langle \psi_2|\psi_1\rangle\}]
\end{eqnarray}
Now it is clearly evident that equations (36) and (38)are not
identical. This indicates that the action of C-Not gate on Bob's
qubit caused the change in the density matrix describing Alice's
subsystem. In other words we can say that local action performed
by Bob on his qubit allowed super luminal signalling to take
place. But in reality, this is not possible. This leads us into a
contradiction and henceforth we conclude that C-Not gate for non
orthogonal set of qubits cannot exist in reality.
\begin{eqnarray}
\end{eqnarray}
\section{Conclusion:}
Here in this work we presented a systematic overview of the
existing impossible operations in quantum domain and their
relationship with no signalling principle. In this work we not
only demonstrate the existing impossibility proofs of various
physical operations but also showed the impossibility of
construction of gates like pauli gates and C-Not gate for
arbitrary qubits from the no signalling principle. As these gates
are the building block for universal quantum gates, one may look
out for the answer that whether the construction of universal
quantum gates for arbitrary qubit is possible from no signalling
principle or not.
\section{Acknowledgement:}
I acknowledge Prof B.S. Choudhury and Satyabrata Adhikari,
Department of Mathematics, Bengal Engineering and Science
University for having useful discussions. I also acknowledge Prof
C.G.Chakraborti, for being the source of inspiration in carrying
out the research.
\section{References:}
$[1]$ W.K.Wootters and W.H.Zurek,Nature 299,802(1982).\\
$[2]$ A.K.Pati and S.L.Braunstein, Nature 404,164(2000).\\
$[3]$ A.K.Pati and Barry C.Sanders, Phys. Lett. A 359, 31-36
(2006).\\
$[4]$ Duanlu Zhou, Bei Zeng, and L. You, Phys. Lett. A 352, 41
(2006).\\
$[5]$ A.K.Pati and S.L.Braunstein, e-print quant-ph/0303124.\\
$[6]$ A.K.Pati, Phys.Rev.A 66, 062319 (2002).\\
$[7]$ Valerio Scarani, Sofyan Iblisdir, Nicolas Gisin, Antonio
Acin, Rev. Mod. Phys. 77, 1225-1256 (2005).\\
$[8]$ A.K.Pati and S.L.Braunstein,Phys.Lett.A 315,208-212
(2003).\\
$[9]$ Indranil Chakrabarty, Satyabrata Adhikari, B.S. Choudhury;
Revisiting impossible quantum operations using principle of
no-signalling and conservation of entanglement under LOCC, e-print
quant-ph/0605186.\\
$[10]$ Indranil Chakrabarty, Satyabrata Adhikari, Prashant and B.S
Choudhury;Inseparability of Quantum Parameters, e-print
quant-ph/0602016.\\
$[11]$ H.P.Yuen, Phys.Lett.A. 113, 405 (1986).\\
$[12]$ I.Chattopadhyay.et.al. Phys. Lett. A, 351, 384-387
(2006).\\
$[13]$ Michal Horodecki, Ryszard Horodecki, Aditi Sen De, Ujjwal
Sen ;No-deleting and no-cloning principles as consequences of
conservation of quantum information , e-print quant-ph /0306044.\\
$[14]$ N. Gisin , Phys . Lett. A 242,1-3 (1998).\\
$[15]$ Indranil Chakrabarty, Satyabrata Adhikari, B.S. Choudhury,
Phys. Scr.74 (2006) 555-557.\\
$[16]$ A .K. Pati Phys. Lett. A 270 103.\\
$[17]$ D. Bruss, G. M. D'Ariano, C. Macchiavello, M. F. Sacchi,
Phys. Rev. A 62, 62302 (2000).\\
$[18]$ Indranil Chakrabarty.et.al, Self replication and Signalling
(communicated).\\
$[19]$ P. Parashar, On the non-existence of a universal Hadamard
gate; e-print quantph/ 0606231.
\end{document}